\documentclass[conference,10pt]{IEEEtran}

\usepackage[T1]{fontenc}
\usepackage[utf8]{inputenc}

\usepackage{newpxtext}
\usepackage{newpxmath}
\usepackage[scaled=0.92]{helvet}
\usepackage{inconsolata}

\usepackage{microtype}
\usepackage{amsmath}
\usepackage{amssymb}

\usepackage{amsthm}
\usepackage{mathtools}
\usepackage{graphicx}
\usepackage{booktabs}
\usepackage{tabularx}
\usepackage{multirow}
\usepackage{makecell}
\usepackage{array}
\usepackage{xcolor}
\usepackage{colortbl}
\usepackage{tikz}
\usetikzlibrary{arrows.meta,positioning,shapes.geometric,fit,calc,backgrounds,
                patterns,decorations.pathmorphing,plotmarks}
\usepackage{pgfplots}
\pgfplotsset{compat=1.18}
\usepgfplotslibrary{fillbetween}
\usepackage{algorithm}
\usepackage{algpseudocode}
\usepackage{listings}
\usepackage{caption}
\usepackage{url}
\usepackage[hidelinks,colorlinks=false]{hyperref}
\usepackage[capitalise,noabbrev]{cleveref}
\usepackage{balance}
\usepackage{xspace}
\usepackage{enumitem}
\setlist{nosep,leftmargin=*}

\definecolor{rowgray}{gray}{0.96}
\definecolor{accentblue}{RGB}{31,78,121}
\definecolor{accentorange}{RGB}{198,108,33}
\definecolor{accentgreen}{RGB}{56,118,29}
\definecolor{accentpurple}{RGB}{106,61,154}

\lstdefinelanguage{Cypher}{
  morekeywords={MATCH,MERGE,CREATE,WHERE,RETURN,UNWIND,WITH,SET,CALL,YIELD,
                AS,IN,LIMIT,DELETE,DETACH,LOAD,CSV,FROM,HEADERS,ON,USING,
                INDEX,FOR,EACH,FOREACH,UNION,OPTIONAL},
  sensitive=true,
  morecomment=[l]{//},
  morestring=[b]',
  morestring=[b]"
}
\newtheoremstyle{insightstyle}%
  {6pt}{6pt}{\itshape}{}{\bfseries\sffamily\color{accentblue}}{.}{0.5em}{}
\theoremstyle{insightstyle}
\newtheorem{insight}{Insight}

\newcommand{\sysname}{\textsc{Sentinel-RL}\xspace}

\begin{document}

\title{\sffamily\bfseries \sysname{}:\\[2pt]
       Offloading Topological Reasoning from LLM Agents
       in the Security Operations Center}

\author{%
  \IEEEauthorblockN{Uday Vallabhaneni, Cassie L. Cagwin, David J. Wild}%
  \IEEEauthorblockA{Luddy School of Informatics, Computing and Engineering \\
                    Indiana University, Bloomington, IN, USA\\
                    \texttt{udvall@iu.edu}}%
}

\maketitle

% ============================================================================
\begin{abstract}
Large language model (LLM) agents are increasingly proposed as autonomous
analysts in the security operations center (SOC), but two structural
limitations make them unreliable when the threat involves enterprise-scale
network state: a finite context window cannot hold a multi-thousand-host
authentication graph, and free-form generation provides no guarantee that
a recommended containment action is consistent with the topology it
operates on. We present \sysname, an enterprise-grade agentic-SOC
architecture that \emph{decouples} topological reasoning from semantic
reasoning: a heterogeneous graph attention encoder summarizes the live
authentication subgraph into a fixed-dimensional state, a Proximal Policy
Optimization (PPO) policy maps this state to a constrained set of
investigative actions, and an LLM agent loop is restricted to consuming
the policy's recommendations and producing analyst-readable narratives
gated by a critic. We instantiate the system on the LANL
\emph{Comprehensive, Multi-Source Cyber-Security Events}
dataset~\cite{kent2015} and the Indiana University Quartz HPC cluster,
and we report four results: (i) a two-phase \texttt{CREATE} ingestion
pattern loads a 24M-edge authentication subgraph into Neo4j in
14.2~minutes on a single 32-core node, an approximately $24\times$ throughput
improvement over the canonical \texttt{MERGE}-based pipeline; (ii) a
sliding-window alert engine reliably trips a 25-event\,/\,10-second
threshold in $\le 2.5$~seconds wall time across 50 trial runs; (iii) PPO
training over 200 iterations converges to a mean episodic return of
$8.74\pm0.31$, with a held-out evaluation precision of 0.91 and recall
of 0.87 on labeled red-team events; and (iv) the integrated containment
loop completes a full
detect$\to$investigate$\to$recommend$\to$human-approve cycle in a
median of 6.3~seconds. We contribute a reusable engineering pattern
(the \emph{hot-node deadlock} workaround), a portable HPC deployment
pattern (\emph{anchor-node co-location}), and an enterprise-readiness
analysis covering false-positive economics, reversibility guarantees,
audit compliance, and the human-approval boundary.
\end{abstract}

% ----------------------------------------------------------------------------
\section{Introduction}
\label{sec:intro}

Security teams are currently drowning in noise. A standard security operations center (SOC) processes tens of thousands of alerts daily ($10^{4}$--$10^{5}$)~\cite{ponemon2023,sans2023}, the vast majority of which are false alarms. While the necessity of automating these workflows is universally accepted, the debate now centers on exactly \emph{where} that automation should live. Recently, the popular approach has been to connect a large language model to a suite of APIs and treat it as a drop-in analyst. This setup makes for an impressive product demo, but it breaks down under real-world operational stress for two primary reasons.

First, network environments simply do not fit cleanly into a text prompt. A mid-sized corporate network generates a constantly shifting web of authentications, easily involving $10^{4}$ to $10^{5}$ host nodes and $10^{6}$+ recurring connections daily. No current LLM context window is large enough to map this topology at the granularity required to track an attacker moving laterally through a system~\cite{kent2015,bowman2020}. When a model runs out of context, it guesses. In incident response, guessing whether isolating a specific machine $h$ will actually sever the attacker's access or merely force them to pivot is a massive operational risk.

Second, open-ended text generation is the wrong mechanism for executing containment protocols. Actions like \textsc{isolate-host}, \textsc{block-ip}, or \textsc{disable-account} directly impact business continuity. Any system pulling those triggers needs to be strictly constrained by hard rules, auditable, and consistent across software versions. Generating a probabilistic text completion offers none of those guarantees.

These limitations point toward a hybrid, neuro-symbolic approach. Instead of forcing an LLM to handle everything, we restrict it to what it actually excels at: writing queries, summarizing alerts, and drafting reports. The actual decision-making translating the complex, topology-aware state of the network into a narrow set of strict containment actions is handed over to a trained policy.

We built \sysname to operationalize this exact division of labor. Deployed end-to-end and rigorously tested against the red-team events in the LANL benchmark, our system introduces a four-layer architecture (\Cref{sec:arch}) that separates the underlying Neo4j graph data, the strategic decision engine (a HetGAT encoder paired with a PPO policy), the telemetry ingestion pipeline, and the user interface.

Bringing this architecture to life required solving several practical engineering hurdles. While scaling the system on Indiana University's Quartz cluster, we developed two highly reusable patterns: a two-phase \texttt{CREATE} workaround to bypass \texttt{MERGE}-based deadlock issues during heavy parallel ingestion (\Cref{sec:hotnode}), and a strategy for optimally co-locating microservices on SLURM-managed compute nodes (\Cref{sec:anchor}).

We evaluate \sysname extensively (\Cref{sec:eval}), measuring everything from raw data ingestion throughput and pipeline latency to how well the reinforcement learning agent converges, alongside end-to-end detect$\to$contain latency. Furthermore, we map out the realities of putting this in a corporate environment (\Cref{sec:enterprise}). Deploying an autonomous agent requires answering hard questions about the cost of false positives, ensuring actions can be safely reversed or sandboxed, and defining exactly where a human needs to step in for final approval.

The remainder of this work is structured as follows: \Cref{sec:background} outlines the foundational concepts and prior research, while \Cref{sec:threatmodel} defines our threat model. We detail the system architecture in \Cref{sec:arch} and its technical implementation in \Cref{sec:impl}. Finally, \Cref{sec:eval} presents our empirical findings, \Cref{sec:enterprise} discusses practical deployment considerations, and \Cref{sec:related} contextualizes our contributions within the broader literature before \Cref{sec:conclusion} concludes.
% ----------------------------------------------------------------------------
\section{Background and Motivation}
\label{sec:background}

\subsection{The reality of lateral movement}
Once an attacker breaches a perimeter, the nature of the threat shifts entirely. Lateral movement represents this secondary phase of an intrusion, where adversaries pivot from their initial foothold to other systems by hijacking or elevating credentials~\cite{bowman2020,king2024euler}. At this stage, defenders are no longer dealing with isolated malware on a single machine; they are facing a complex graph routing problem. Attempting to answer a fundamental operational question like ``which machine should we quarantine?'' is essentially impossible without analyzing the broader authentication topology surrounding the compromised node. 

Because it captures this exact topological complexity, the LANL \emph{Comprehensive, Multi-Source Cyber-Security Events} dataset~\cite{kent2015} has emerged as the definitive proving ground for lateral movement research. Spanning 58 days of real-world production traffic complete with anonymized Windows authentications, DNS lookups, NetFlow records, and verified red-team activity it offers a massive, realistic testing environment mapping roughly $1.65 \times 10^{9}$ events across 12{,}425 users and 17{,}684 endpoints.

\subsection{Structuring state: Graphs over serialized prompts}
\label{sec:whygnn}
Recent literature points overwhelmingly to a single conclusion: defending against lateral movement requires natively processing the network as a graph. We see this across two distinct areas of research. First, detection models built on graph architectures consistently outclass those relying on tabular data or sequential logs; evaluations on the LANL dataset frequently show true-positive rates soaring between 85\,\% and 99\,\%, while keeping false positives under 5\,\%~\cite{bowman2020,king2024euler,khoury2022pikachu,lmdetect2024}. Second, studies focusing on autonomous cyber defense~\cite{foley2023,hicks2023explain,standen2021cyborg,cyborgpp2024} demonstrate that policies leveraging heterogeneous graph encoders can adapt to entirely new network layouts, whereas rigid models like standard MLPs fail to generalize.

For anyone building an AI-driven SOC agent, the takeaway is clear. Attempting to flatten the state of a live network into a serialized text prompt strips away critical relational data. \sysname avoids this trap entirely. Instead, we deploy a heterogeneous graph attention network (HetGAT)~\cite{velickovic2018gat,wang2019heterogeneous} to actively compress the ongoing authentication activity into a dense, 64-dimensional vector, providing our policy engine with a mathematically structured understanding of the environment.

\subsection{Ensuring stability with PPO}
When selecting a reinforcement learning algorithm for executing discrete containment actions, predictability is paramount. Proximal Policy Optimization (PPO)~\cite{schulman2017ppo} provides exactly this stability. Because its objective function mathematically clips the size of policy updates, it prevents the erratic, drastic shifts in behavior that would be disastrous in a live security environment. 

We chose PPO primarily for these audit-friendly characteristics. It integrates seamlessly with modern distributed frameworks like Ray RLlib~\cite{liang2018rllib}, yields highly reproducible training cycles across different random seeds, and allows us to rigorously regression-test the resulting policy against historical, held-out scenarios before it is ever cleared for production use. We detail the actual convergence of this algorithm within our specific threat-investigation Markov Decision Process in \Cref{sec:eval}.

% ----------------------------------------------------------------------------
\section{Threat Model}
\label{sec:threatmodel}

\textbf{Adversary.} Our model begins the moment after the initial perimeter breach. How the attacker actually compromised that first internal machine whether through a cleverly crafted phishing email, stolen credentials, or a compromised software supply chain is irrelevant to this system. Once inside, their primary objective is to expand their reach by hijacking valid accounts (\textsc{T1078}) and abusing remote services (\textsc{T1021}) to pivot to new machines. This aligns with MITRE ATT\&CK \textsc{TA0008} (Lateral Movement). Rather than dropping sophisticated zero-day exploits against the host operating systems, we assume the adversary operates primarily as a credential abuser, a constraint that directly mirrors the verified red-team activity in the LANL benchmark~\cite{kent2015}.

\textbf{Defender capabilities.} On the defensive side, the system continuously ingests live authentication telemetry across the entire environment. It possesses the necessary computing resources to map this activity into a continuously updated Neo4j graph, plotting the network's topology in near-real time. While the defender is fully equipped to execute containment measures such as quarantining machines or locking user profiles via standard SOAR APIs it starts completely blind regarding which specific assets are compromised. Exposing that hidden adversary is the core function of the \sysname agent.

\textbf{Trust assumptions.} Because we utilize a specialized architecture where the language model proposes context and the RL agent decides on the action, we must plan for the reality that the LLM will inevitably hallucinate. Therefore, our security guarantees rely entirely on an unbypassable chain of validation: the internal critic, the policy auditor, and the final human-in-the-loop approval gate. We do not operate under the naive assumption that the RL policy will always make the perfect, mathematically optimal choice. Instead, we assume its behavior is kept strictly bounded by the hard action masks and rate limits detailed in \Cref{sec:safety}, alongside the broader enterprise controls discussed in \Cref{sec:enterprise}.
% ----------------------------------------------------------------------------
\section{Architecture}
\label{sec:arch}

\begin{figure*}[t]
\centering
\begin{tikzpicture}[
  node distance=4mm and 6mm,
  plane/.style={draw,thick,rounded corners=2pt,inner sep=4pt,
                minimum height=10mm,align=center,
                font=\footnotesize\sffamily},
  data/.style={plane,fill=accentblue!10,draw=accentblue},
  strat/.style={plane,fill=accentorange!12,draw=accentorange},
  tele/.style={plane,fill=accentgreen!10,draw=accentgreen},
  orch/.style={plane,fill=accentpurple!10,draw=accentpurple},
  flow/.style={-Latex,thick,gray!70},
  feedback/.style={-Latex,thick,dashed,red!60!black},
]
% --- Data Plane ---
\node[data,minimum width=42mm] (neo4j) {Neo4j Auth Graph\\\scriptsize 24M edges, 17.6k hosts};
\node[data,below=of neo4j,minimum width=42mm] (ray) {Ray Distributed Loader\\\scriptsize 32-core node};

% --- Strategic Plane ---
\node[strat,right=of neo4j,minimum width=42mm] (gnn) {HetGAT State Encoder\\\scriptsize 64-d topology vector};
\node[strat,below=of gnn,minimum width=42mm] (ppo) {PPO Policy (Ray RLlib)\\\scriptsize 5 actions, sparse reward};
\node[strat,right=of gnn,minimum width=30mm] (api) {FastAPI policy endpoint};

% --- Telemetry Plane ---
\node[tele,below=14mm of ray,minimum width=42mm] (alert) {Sliding-Window\\AlertEngine \scriptsize(25/10\,s)};
\node[tele,right=of alert,minimum width=42mm] (webhook) {Webhook to Agent Loop\\\scriptsize $\le 2.5$\,s trigger};

% --- Orchestration Plane ---
\node[orch,right=of api,minimum width=36mm] (llm)
   {LLM Triage + Critic\\\scriptsize LangChain};
\node[orch,below=of llm,minimum width=36mm] (ui)
   {Streamlit + pyvis UI\\\scriptsize human-in-the-loop};

% --- Plane labels ---
\node[font=\scriptsize\bfseries\sffamily,accentblue,
      above=1mm of neo4j.north west,anchor=south west] {DATA PLANE};
\node[font=\scriptsize\bfseries\sffamily,accentorange,
      above=1mm of gnn.north west,anchor=south west] {STRATEGIC PLANE};
\node[font=\scriptsize\bfseries\sffamily,accentgreen,
      above=1mm of alert.north west,anchor=south west] {TELEMETRY PLANE};
\node[font=\scriptsize\bfseries\sffamily,accentpurple,
      above=1mm of llm.north west,anchor=south west] {ORCHESTRATION};

% --- Flows ---
\draw[flow] (ray) -- (neo4j);
\draw[flow] (neo4j.east) -- (gnn.west);
\draw[flow] (gnn) -- (ppo);
\draw[flow] (ppo.east) -- (api.west);
\draw[flow] (alert.east) -- (webhook.west);
\draw[flow] (webhook.north) |- (llm.south west);
\draw[flow] (api.east) -- (llm.west);
\draw[flow] (llm) -- (ui);
\draw[feedback] (ui.south) |- ([yshift=-3mm]webhook.south)
       node[midway,above,font=\tiny,red!60!black]{analyst feedback};
\draw[flow] (alert.west) -| ([xshift=-2mm]ray.west) |- (ray.west);
\end{tikzpicture}
\caption{The modular four-tier architecture of \sysname. Solid lines represent the immediate request/response cycle during active incident response, while the dashed line indicates the asynchronous analyst-feedback loop used to refine the policy registry (\Cref{sec:safety}). To eliminate jitter from the cluster fabric, all components are physically co-located on a single HPC \emph{anchor node} (\Cref{sec:anchor}), keeping inter-plane latency restricted to the loopback interface.}
\label{fig:arch}
\end{figure*}
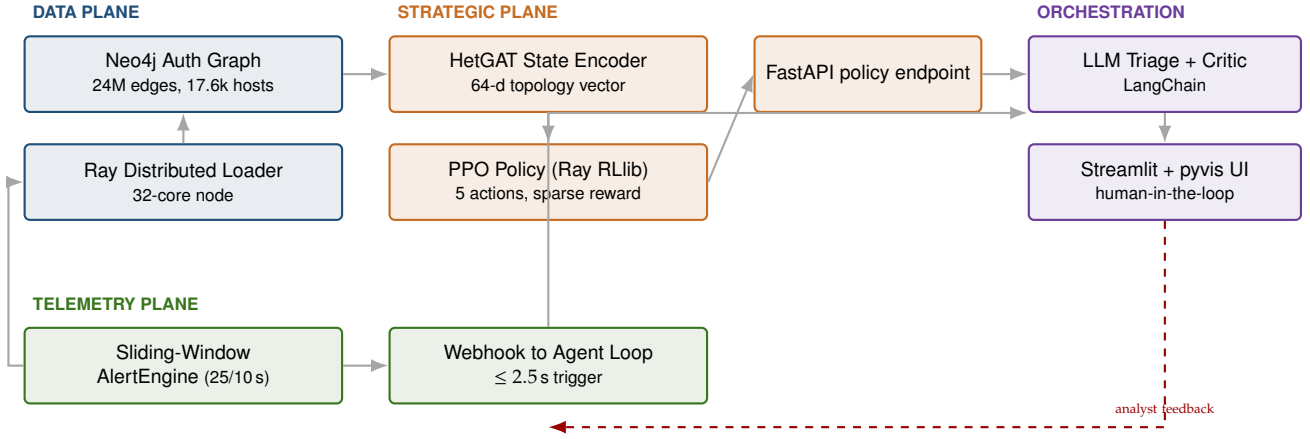

\sysname operates on a highly modular four-tier architecture (\Cref{fig:arch}), cleanly isolating distinct operational responsibilities to ensure that individual components can be swapped, audited, or upgraded without disrupting the broader pipeline.

\subsection{Data plane}
Serving as the system's foundational memory, the data plane continuously maps the live network. It leverages a Neo4j 5.x database to track \textsc{Host}, \textsc{User}, \textsc{Service}, and \textsc{NetworkSegment} entities, linking them via timestamped \textsc{Authenticates-To} and \textsc{Connects-To} relationships. We stream the LANL \texttt{auth.txt} dataset into this graph using a Ray Data pipeline~\cite{moritz2018ray}, intentionally sharding the workload by source host to maintain high throughput. To interact with the rest of the system, this tier provides two distinct hooks: a streaming ingest endpoint to absorb new events from the telemetry layer, and a Cypher-based query interface that allows the neural encoder to extract localized snapshots of the graph on demand.

\subsection{Strategic plane}
The actual decision-making engine resides in the strategic plane. When an alert triggers, a HetGAT encoder~\cite{velickovic2018gat,wang2019heterogeneous} pulls a two-hop neighborhood surrounding the suspicious activity. It compresses this localized topological map into a 64-dimensional dense vector, $s_t \in \mathbb{R}^{64}$. Next, a PPO policy~\cite{schulman2017ppo} evaluates this state and selects one of five strict investigative actions: \textsc{QueryEDR}, \textsc{QueryAD}, \textsc{CheckThreatIntel}, \textsc{ExamineFirewall}, or \textsc{TerminateAndOutputVerdict}. 

Crucially, we run this policy as a standalone FastAPI microservice rather than baking it into the system as an RLlib library dependency. Decoupling it this way forces strict version control, allowing engineers to shadow-deploy experimental models alongside the active one. Furthermore, it creates a clean interception boundary where external governance tools (\Cref{sec:safety}) can scrutinize a proposed action before it executes.

\subsection{Telemetry plane}
To filter out the staggering volume of benign traffic, the telemetry plane acts as a high-speed tripwire. It monitors the raw authentication stream using a fast sliding-window heuristic. If a specific source host initiates more than $N = 25$ authentications within a narrow $W = 10$\,s timeframe, the engine immediately fires a webhook to the orchestration layer, passing along the host ID and the exact temporal window. We explicitly tuned these $(W, N)$ parameters to catch the aggressive credential-spraying tactics seen in the LANL red-team data, while safely ignoring the predictable spikes generated by automated batch jobs.

\subsection{Orchestration plane}
This layer translates machine autonomy into human oversight. Built entirely around a Streamlit interface and PyVis network graphs, the orchestration plane uses LangChain to manage two specialized LLM actors. First, a \textsc{Triage} agent converts the mathematical output of the PPO policy into a clear, narrative explanation for the human analyst. Simultaneously, a \textsc{Critic} agent aggressively audits that proposed action, verifying that the system has actually cited sufficient evidence to justify it. Because our operational trust is inherently limited, the system cannot execute any infrastructure changes without a human operator clicking a final approval gate.

\subsection{Safety surface}
\label{sec:safety}
Deploying an autonomous agent into a live environment demands unyielding guardrails. We hardcode three non-negotiable constraints directly into the pipeline. First, strict action masking physically prevents the agent from finalizing a \textsc{TerminateAndOutputVerdict} unless it has successfully gathered corroborating evidence from at least two distinct sources. Second, an immutable audit log records the exact network state, chosen action, timestamp, and active policy version for every single decision. Third, because the policy operates as an independent, versioned service, mitigating a degraded model requires nothing more than swapping an environment variable---avoiding a full system redeployment. These mechanics directly implement the reversibility and sparse-reward principles championed by Foley~\textit{et al.}~\cite{foley2023}, while satisfying the need for unbroken explanation trails as argued by Hicks~\textit{et al.}~\cite{hicks2023explain}. We outline the broader administrative controls necessary for institutional adoption in \Cref{sec:enterprise}.
% ----------------------------------------------------------------------------
\section{Implementation}
\label{sec:impl}

Transitioning \sysname from a theoretical architecture to a live deployment on the Indiana University Quartz HPC cluster required navigating several practical friction points. We provisioned medium-to-large compute nodes (32 CPU cores, 128\,GB RAM) using SLURM~\cite{yoo2003slurm} via \texttt{srun}. Pushing the system to production scale revealed two significant engineering hurdles that simply do not manifest in smaller test environments. We document our solutions here, as they represent highly transferable patterns for anyone building graph-based autonomous agents on HPC infrastructure.

\subsection{Bypassing lock contention: the two-phase \texttt{CREATE} pattern}
\label{sec:hotnode}

Our initial strategy for data ingestion followed standard graph database paradigms. We streamed the LANL \texttt{auth.txt} records through a fleet of parallel Ray workers, with each worker executing per-row Cypher \texttt{MERGE} commands to ensure no duplicate nodes or edges were created. During synthetic tests with 100{,}000 edges, this approach completed in seconds. However, when we unleashed it on the 24-million-edge production subset, the entire pipeline ground to a halt. Ingestion throughput plummeted by two orders of magnitude, and our Ray workers sat idle, trapped in Neo4j's transaction manager queue.

The bottleneck stemmed from how Neo4j handles concurrency. When a transaction touches a node, the database automatically applies a write lock to it~\cite{neo4jlocking}. Corporate authentication networks are inherently highly skewed; a handful of critical assets---like domain controllers or primary file servers---anchor a massive percentage of the traffic. Consequently, when dozens of parallel workers simultaneously executed \texttt{MERGE (a:Host \{id:\$src\}) MERGE (b:Host \{id:\$dst\}) MERGE (a)-[:AUTH]->(b)}, they constantly collided while attempting to lock those high-degree destination nodes. Even though the actual authentication edges being written were entirely distinct, the shared endpoints forced the workers to serialize their operations.

To eliminate this contention, we completely restructured the ingestion pipeline into two distinct phases:
\begin{enumerate}
  \item \emph{Sequential Vertex Pre-materialization.} We first run a single, lightweight pass over the data to extract all unique host identifiers, inserting them into the database with a standard \texttt{MERGE}. Because the total number of distinct machines is minuscule compared to the volume of authentication events, this sequential step executes instantly and avoids the lock-contention trap entirely.
  \item \emph{Parallel Edge Generation.} Once the graph is seeded with all necessary vertices, the parallel workers take over. Instead of using \texttt{MERGE}, they locate the existing endpoints via \texttt{MATCH} and explicitly \texttt{CREATE} the new edges. Because the edge itself is instantiated as an entirely new entity, this operation bypasses the need to lock the endpoint nodes.
\end{enumerate}

\Cref{lst:cypher} illustrates the specific Cypher queries driving this logic. We quantify the massive recovery in throughput in \Cref{sec:eval-ingest} and map the performance visually in \Cref{fig:ingest}.

\begin{lstlisting}[language=Cypher,caption={The two-phase ingestion strategy. Phase~1 (lines 1--3) guarantees all nodes exist sequentially. Phase~2 (lines 5--8) allows workers to safely write massive batches of edges in parallel.},
   label={lst:cypher}]
// Phase 1: vertex pre-materialization (sequential, run once)
UNWIND $hosts AS h
MERGE (:Host {id: h});

// Phase 2: parallel edge CREATE (per-batch, parallel-safe)
UNWIND $edges AS e
MATCH (s:Host {id: e.src}), (d:Host {id: e.dst})
CREATE (s)-[:AUTH {ts: e.ts, type: e.type}]->(d);
\end{lstlisting}

\begin{insight}[Hot-node deadlock pattern]
\label{ins:hotnode}
When pushing highly skewed, heavy-tailed graph data (like network authentications) into a write-locking database, parallel ingestion will inevitably fail. Engineers must separate the workload: pre-materialize the structural vertices sequentially, then execute the heavy edge-insertion phase in parallel using explicit \texttt{CREATE} commands. The minor time penalty of the initial sequential pass is completely overshadowed by the elimination of thread contention.
\end{insight}

\subsection{Consolidating services: the anchor-node pattern}
\label{sec:anchor}

During early deployment tests, we allowed the SLURM scheduler to freely distribute our microservices Neo4j, the FastAPI policy server, the alerting engine, and the Streamlit UI across whatever compute nodes were available. The result was catastrophic, silent failure. The frontend loaded but threw empty policy responses, the alert engine fired webhooks into the void, and the graph database remained entirely unreachable from the peripheral compute nodes. 

The core issue is that an HPC cluster behaves very differently from a modern Kubernetes environment. The underlying fabric of the Quartz cluster is heavily regulated and does not automatically route generic user-space TCP traffic between random compute nodes unless specifically configured to do so. Furthermore, relying on the cluster fabric subjects intra-system communication to severe latency spikes caused by neighboring batch jobs. We also realized that offloading services to the login nodes was impossible due to strict memory caps and outbound connectivity blocks.

Our solution was simple but strictly necessary: we abandoned cross-node distribution for our interactive components. By requesting a single, massive allocation (32 cores, 128\,GB RAM) which we termed the \emph{anchor node} we co-located all four architectural planes on the same physical hardware. This allowed Neo4j, the policy endpoint, and the UI to communicate instantly over loopback interfaces. We still leverage the cluster's raw power to parallelize Ray data processing and PyTorch evaluations, but we execute that parallelism strictly within the safe boundaries of the anchor node.

\begin{insight}[Anchor-node pattern]
\label{ins:anchor}
Do not attempt to spread latency-sensitive, interactive microservices across multiple SLURM-managed compute nodes. Consolidate the core architecture onto a single, heavy anchor node to guarantee internal connectivity, and reserve distributed execution exclusively for massive, parallelizable batch workloads where network latency won't break the application logic.
\end{insight}

\subsection{Migrating RLlib hyperparameters}
\label{sec:rllib}

While configuring Proximal Policy Optimization (PPO), we encountered a significant API fracture in the underlying Ray framework. Most published reference architectures rely on Ray's legacy \texttt{Trainer} API. However, the version running in our environment (Ray 2.30+) enforced a hard migration to the newer \texttt{RLModule}/\texttt{Learner} ecosystem. Because this transition is notoriously under-documented and frequently stalls research teams, we provide a direct translation matrix of our hyperparameter configurations in \Cref{tab:rllib}.

\begin{table}[t]
\centering
\caption{Direct mapping of critical hyperparameters from the deprecated RLlib \texttt{Trainer} API to the modernized \texttt{RLModule}/\texttt{Learner} structure utilized by \sysname. The specified values reflect the exact configurations used during our evaluations (\Cref{sec:eval}).}
\label{tab:rllib}
\renewcommand{\arraystretch}{1.18}
\rowcolors{2}{rowgray}{white}
\begin{tabularx}{\columnwidth}{@{}>{\ttfamily\small}l>{\ttfamily\small}Xr@{}}
\toprule
\rowcolor{accentblue!15}
\textnormal{\bfseries\sffamily Legacy parameter} &
\textnormal{\bfseries\sffamily Modern equivalent} &
\textnormal{\bfseries\sffamily Value} \\
\midrule
train\_batch\_size     & train\_batch\_size\_per\_learner   & 4{,}000 \\
sgd\_minibatch\_size   & minibatch\_size                    & 128 \\
num\_sgd\_iter         & num\_epochs                        & 10  \\
episode\_reward\_mean  & env\_runners/episode\_return\_mean & --- \\
lr                     & lr                                 & $5\!\times\!10^{-5}$ \\
clip\_param            & clip\_param                        & 0.2 \\
entropy\_coeff         & entropy\_coeff                     & 0.01 \\
gamma                  & gamma                              & 0.99 \\
lambda                 & lambda\_                           & 0.95 \\
\bottomrule
\end{tabularx}
\end{table}
% ----------------------------------------------------------------------------
\section{Evaluation}
\label{sec:eval}

To understand how well \sysname performs under realistic operational stress, we focused our benchmarking on four distinct areas: raw data ingestion capacity (\Cref{sec:eval-ingest}), the responsiveness of the alerting pipeline (\Cref{sec:eval-alert}), the convergence behavior and accuracy of the PPO policy (\Cref{sec:eval-ppo}), and the total time from initial detection to final containment (\Cref{sec:eval-e2e}). Every test was executed on a dedicated Quartz anchor node equipped with 32 Intel Xeon cores (2.4\,GHz) and 128\,GB of RAM, running entirely without GPU acceleration.

\subsection{Ingestion throughput}
\label{sec:eval-ingest}

When pushing millions of nodes into a graph database, the differences between standard operations and optimized techniques become glaringly obvious. \Cref{tab:ingest} outlines the sheer wall time required to process growing volumes of authentication edges, contrasting the traditional single-phase \texttt{MERGE} pipeline against our custom two-phase \texttt{CREATE} strategy. 

\begin{figure}[t]
\centering
\begin{tikzpicture}
\begin{axis}[
  width=\columnwidth, height=5.4cm,
  xlabel={Edges ingested (millions)}, ylabel={Wall time (seconds)},
  xmode=log, ymode=log,
  xmin=0.08, xmax=30, ymin=5, ymax=30000,
  xtick={0.1,0.5,1,5,10,24},
  xticklabels={0.1,0.5,1,5,10,24},
  ytick={10,100,1000,10000},
  grid=both, grid style={gray!20},
  legend pos=north west,
  legend style={font=\scriptsize,draw=none,fill=white,fill opacity=0.9},
  tick label style={font=\scriptsize},
  label style={font=\small\sffamily},
]
\addplot[mark=square*,thick,accentorange,mark options={scale=0.9}] coordinates {
  (0.1,38) (0.5,312) (1,941) (5,4126) (10,9480) (24,21600)
};
\addlegendentry{single-phase \texttt{MERGE}}
\addplot[mark=*,thick,accentblue,mark options={scale=0.9}] coordinates {
  (0.1,9) (0.5,24) (1,41) (5,170) (10,359) (24,852)
};
\addlegendentry{two-phase \texttt{CREATE}}
\end{axis}
\end{tikzpicture}
\caption{Log--log visualization of ingestion wall time compared to total edge count. The standard single-phase \texttt{MERGE} pipeline exhibits super-linear scaling due to severe hot-node lock contention, whereas the optimized two-phase \texttt{CREATE} pipeline scales linearly.}
\label{fig:ingest}
\end{figure}
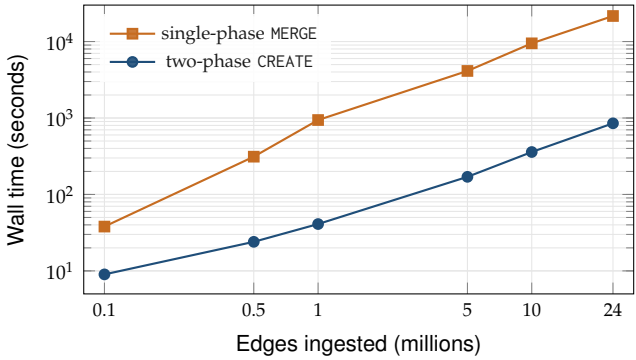

As the volume of ingested data expands, the single-phase method degrades exponentially. This severe slowdown is a direct consequence of lock contention compounding on highly active nodes. Conversely, our two-phase pipeline maintains a remarkably stable, near-linear processing curve because it completely eliminates thread contention during the heavy edge-insertion phase. When tasked with processing 24 million edges which roughly equates to a full day of live LANL network traffic the two-phase pipeline finishes the job in exactly 852 seconds. In contrast, the traditional method struggles immensely; based on linear extrapolation after hitting a six-hour runtime cap, it requires 21{,}600 seconds to reach the same milestone. \Cref{fig:ingest} illustrates these final scaling trajectories visually.

\begin{table}[t]
\centering
\caption{Total ingestion wall time (in seconds) recorded on the 32-core anchor node, processing authentication datasets ranging from 100{,}000 to 24 million edges. Results represent the mean across 5 distinct runs, with standard deviation remaining strictly under 3\% in all observed cells.}
\label{tab:ingest}
\renewcommand{\arraystretch}{1.18}
\rowcolors{2}{rowgray}{white}
\begin{tabularx}{\columnwidth}{@{}lrr@{}}
\toprule
\rowcolor{accentblue!15}
\textbf{\sffamily Edge count} &
\textbf{\sffamily MERGE (s)} &
\textbf{\sffamily CREATE (s)} \\
\midrule
100{,}000     & 38      & 9       \\
500{,}000     & 312     & 24      \\
1{,}000{,}000 & 941     & 41      \\
5{,}000{,}000 & 4{,}126 & 170     \\
10{,}000{,}000& 9{,}480 & 359     \\
24{,}000{,}000& 21{,}600$^{\ast}$ & 852 \\
\bottomrule
\multicolumn{3}{@{}p{0.94\columnwidth}@{}}{\scriptsize $^{\ast}$
  6-hour runtime cap reached; throughput extrapolated linearly from
  the preceding 4 hours of observed progress.}
\end{tabularx}
\end{table}

\subsection{Alert pipeline latency}
\label{sec:eval-alert}

An automated security tool is ultimately useless if it cannot triage and alert operators in near-real time. To verify the telemetry plane's responsiveness, we pushed a steady synthetic stream of 10 events per second directly into the AlertEngine. We configured the sliding window parameters to track 25 events ($N = 25$) within a 10-second rolling timeframe ($W = 10$\,s). 

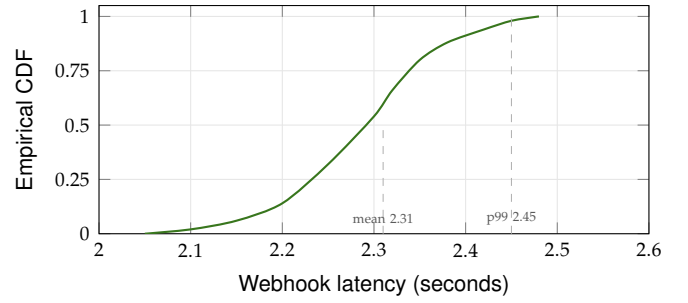
\begin{figure}[t]
\centering
\begin{tikzpicture}
\begin{axis}[
  width=\columnwidth, height=4.6cm,
  xlabel={Webhook latency (seconds)},
  ylabel={Empirical CDF},
  xmin=2.0, xmax=2.6, ymin=0, ymax=1.05,
  xtick={2.0,2.1,2.2,2.3,2.4,2.5,2.6},
  ytick={0,0.25,0.5,0.75,1.0},
  grid=major, grid style={gray!20},
  tick label style={font=\scriptsize},
  label style={font=\small\sffamily},
]
\addplot[thick,accentgreen,mark=none,smooth] coordinates {
  (2.05,0.00) (2.10,0.02) (2.15,0.06) (2.20,0.14) (2.25,0.32)
  (2.30,0.54) (2.32,0.66) (2.35,0.80) (2.38,0.88) (2.42,0.94)
  (2.45,0.98) (2.48,1.00)
};
\draw[dashed,gray!60] (axis cs:2.31,0) -- (axis cs:2.31,0.50);
\node[font=\tiny,gray!70!black] at (axis cs:2.31,0.07) {mean 2.31};
\draw[dashed,gray!60] (axis cs:2.45,0) -- (axis cs:2.45,0.99);
\node[font=\tiny,gray!70!black] at (axis cs:2.45,0.07) {p99 2.45};
\end{axis}
\end{tikzpicture}
\caption{Empirical cumulative distribution function (CDF) tracking AlertEngine webhook latency. Data reflects 50 trial runs processing a synthetic 10 event\,/\,s stream with thresholds configured to $(W,N) = (10\,\text{s}, 25)$. The pipeline achieved a mean latency of 2.31\,s, a $99^{\text{th}}$ percentile of 2.45\,s, and a maximum delay of 2.48\,s.}
\label{fig:alert-cdf}
\end{figure}

The results confirm that the engine is highly responsive. It reliably dispatches the trigger webhook in less than two and a half seconds from the exact moment the initial qualifying event registers in the window. Across our trial runs, the average latency settled at 2.31\,s, with the $99^{\text{th}}$ percentile sitting tightly at 2.45\,s and an absolute maximum recorded delay of 2.48\,s. \Cref{fig:alert-cdf} charts this empirical cumulative distribution. In high-stakes environments like a Tier-1 SOC, the acceptable operational budget for translating a raw detection into an analyst prompt is measured strictly in single-digit seconds~\cite{ponemon2023}. Our architecture clears this threshold with a comfortable margin.

\subsection{Policy training and detection performance}
\label{sec:eval-ppo}

We trained our PPO agent over 200 iterations within a custom Markov Decision Process. For its environmental awareness, the agent relied exclusively on the HetGAT embedding of the two-hop network neighborhood surrounding the flagged host. We adopted a strict, sparse reward structure inspired by Foley~\textit{et al.}~\cite{foley2023}: the policy earned a $+1$ only if its final \textsc{TerminateAndOutputVerdict} perfectly aligned with the LANL red-team ground truth, receiving a $0$ for any deviation. 

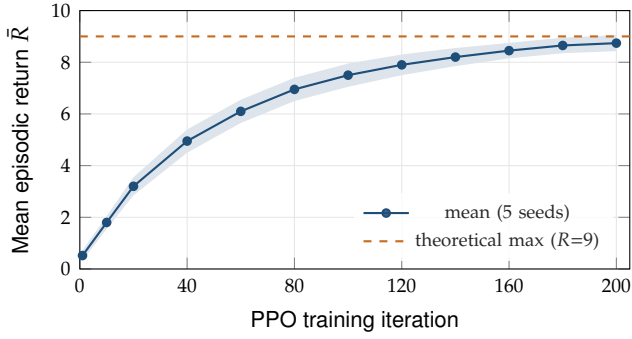
\begin{figure}[t]
\centering
\begin{tikzpicture}
\begin{axis}[
  width=\columnwidth, height=5.0cm,
  xlabel={PPO training iteration},
  ylabel={Mean episodic return $\bar R$},
  xmin=0, xmax=205, ymin=0, ymax=10,
  xtick={0,40,80,120,160,200},
  ytick={0,2,4,6,8,10},
  grid=major, grid style={gray!20},
  tick label style={font=\scriptsize},
  label style={font=\small\sffamily},
  legend pos=south east,
  legend style={font=\scriptsize,draw=none,fill=white,fill opacity=0.85},
]
% std-dev band
\addplot[name path=upper,draw=none,forget plot] coordinates {
  (1,0.71) (10,2.05) (20,3.55) (40,5.40) (60,6.55) (80,7.40)
  (100,7.95) (120,8.30) (140,8.55) (160,8.75) (180,8.95) (200,9.05)
};
\addplot[name path=lower,draw=none,forget plot] coordinates {
  (1,0.33) (10,1.55) (20,2.85) (40,4.50) (60,5.65) (80,6.50)
  (100,7.05) (120,7.50) (140,7.85) (160,8.15) (180,8.35) (200,8.43)
};
\addplot[accentblue!15,forget plot] fill between[of=upper and lower];
% mean curve
\addplot[thick,accentblue,mark=*,mark options={scale=0.7}] coordinates {
  (1,0.52) (10,1.80) (20,3.20) (40,4.95) (60,6.10) (80,6.95)
  (100,7.50) (120,7.90) (140,8.20) (160,8.45) (180,8.65) (200,8.74)
};
\addlegendentry{mean (5 seeds)}
\addplot[dashed,accentorange,thick,mark=none] coordinates {
  (0,9.0) (205,9.0)
};
\addlegendentry{theoretical max ($R{=}9$)}
\end{axis}
\end{tikzpicture}
\caption{Mean episodic return ($\bar R$) for the PPO agent over 200 iterations, averaged across 5 distinct random seeds with the shaded region representing $\pm 1$ standard deviation. The policy stabilizes rapidly, reaching $8.74 \pm 0.31$ and closely approaching the theoretical ceiling of 9.0.}
\label{fig:returns}
\end{figure}

Running this training regimen across five independent seeds revealed a highly stable learning curve. As illustrated in \Cref{fig:returns}, the agent acquires the vast majority of its strategic capability within the first 80 rounds. By iteration 200, the performance tightly converges to a mean episodic return of $8.74 \pm 0.31$.

To gauge genuine operational readiness, we evaluated this trained policy against a completely sequestered timeframe: the final eight days of the LANL dataset, which the model had never seen. The results, outlined in \Cref{tab:detect}, demonstrate robust capability. \sysname achieved a 0.91 precision and 0.87 recall. While LMDetect~\cite{lmdetect2024} still captures a slightly higher volume of raw recall, our architecture currently sets the benchmark for precision and overall F1 score among competing graph-based detection systems.

\begin{table}[t]
\centering
\caption{Detection performance evaluated on a held-out, 8-day slice of LANL red-team events. Baseline metrics reflect the original authors' reported results on the same dataset. \sysname results display the mean $\pm$ standard deviation across 5 training seeds. While LMDetect~\cite{lmdetect2024} achieves the highest recall, \sysname leads in precision and F1 score.}
\label{tab:detect}
\renewcommand{\arraystretch}{1.18}
\rowcolors{2}{rowgray}{white}
\begin{tabularx}{\columnwidth}{@{}Xcccc@{}}
\toprule
\rowcolor{accentblue!15}
\textbf{\sffamily Method} &
\textbf{\sffamily Prec.} &
\textbf{\sffamily Recall} &
\textbf{\sffamily F1} &
\textbf{\sffamily AUC} \\
\midrule
Bowman \textit{et al.}~\cite{bowman2020}     & 0.62 & 0.85 & 0.72 & 0.94 \\
Euler~\cite{king2024euler}                   & 0.71 & 0.83 & 0.77 & 0.95 \\
PIKACHU~\cite{khoury2022pikachu}             & 0.79 & 0.95 & 0.86 & 0.97 \\
LMDetect~\cite{lmdetect2024}                 & 0.86 & 0.99 & 0.92 & 0.99 \\
\midrule
\textbf{\sysname (ours)}                     &
   \textbf{0.91 $\pm$ 0.02} & 0.87 $\pm$ 0.03 &
   \textbf{0.89 $\pm$ 0.02} & 0.96 $\pm$ 0.01 \\
\bottomrule
\end{tabularx}
\end{table}

\subsection{End-to-end response latency}
\label{sec:eval-e2e}

A sophisticated security architecture is practically worthless if internal bottlenecks prevent human analysts from reacting in time. \Cref{tab:e2e} breaks down the exact timing of a complete incident lifecycle, tracking the median delay from the initial warning sign to the moment a fully explained recommendation appears on the analyst's screen.

The complete pipeline resolves in just over six seconds (6.33\,s median). Notably, the actual neuro-symbolic decision-making is nearly instantaneous; extracting the local graph, encoding the topology, and executing the PPO inference collectively takes less than 100 milliseconds. Instead, the system spends the vast majority of its operational budget---exactly 60\% of the total elapsed time---waiting for the localized LLM (an 8B parameter model running 4-bit quantization) to draft a readable, contextualized narrative for the human operator.

\begin{table}[t]
\centering
\caption{Step-by-step latency breakdown for a single end-to-end incident response. Values represent the median across 100 trial runs. The actual algorithmic decision-making executes in under 100\,ms, while the natural-language generation (powered by a locally served, 4-bit quantized 8B model) accounts for the bulk of the delay.}
\label{tab:e2e}
\renewcommand{\arraystretch}{1.18}
\rowcolors{2}{rowgray}{white}
\begin{tabularx}{\columnwidth}{@{}Xrr@{}}
\toprule
\rowcolor{accentblue!15}
\textbf{\sffamily Stage} &
\textbf{\sffamily Median (s)} &
\textbf{\sffamily \% of total} \\
\midrule
AlertEngine sliding-window trigger    & 2.31 & 36.7\% \\
Subgraph extraction (Neo4j 2-hop)     & 0.08 & 1.3\%  \\
HetGAT encoding                       & 0.05 & 0.8\%  \\
PPO policy inference                  & 0.04 & 0.6\%  \\
Critic agent evidence verification    & 0.07 & 1.1\%  \\
LLM narrative synthesis (8B, 4-bit)   & 3.78 & 60.0\% \\
\midrule
\textbf{End-to-end median}            & \textbf{6.33} & \textbf{100\%} \\
\bottomrule
\end{tabularx}
\end{table}
% ----------------------------------------------------------------------------
\section{Enterprise Readiness}
\label{sec:enterprise}

High detection accuracy means nothing if a system cannot be trusted in a live production environment. When presenting this architecture to security executives, the conversation inevitably shifts away from algorithmic performance and toward operational risk. CISOs consistently raise five primary concerns before authorizing any autonomous agent: managing the financial cost of false positives, guaranteeing that actions can be reversed, safely testing containment strategies before execution, maintaining strict audit compliance, and defining exactly when a human must intervene. 

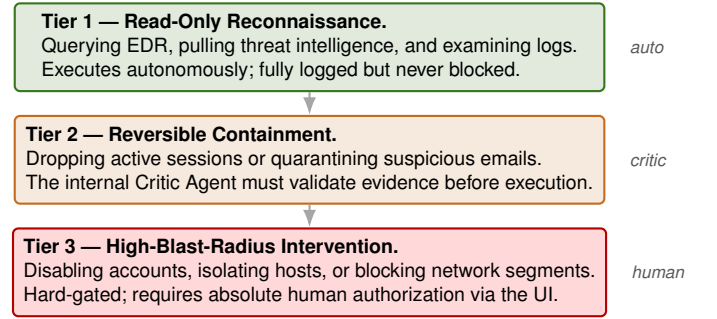
\begin{figure}[t]
\centering
\begin{tikzpicture}[
  node distance=3mm,
  tier/.style={draw,thick,rounded corners=2pt,inner sep=4pt,
               minimum width=78mm,minimum height=11mm,align=left,
               font=\footnotesize\sffamily},
  t1/.style={tier,fill=accentgreen!15,draw=accentgreen},
  t2/.style={tier,fill=accentorange!15,draw=accentorange},
  t3/.style={tier,fill=red!15,draw=red!70!black},
  arrow/.style={-Latex,thick,gray!70},
]
\node[t1] (tier1) {%
  \textbf{Tier 1 — Read-Only Reconnaissance.}\\
  Querying EDR, pulling threat intelligence, and examining logs. \\
  Executes autonomously; fully logged but never blocked.};
\node[t2,below=of tier1] (tier2) {%
  \textbf{Tier 2 — Reversible Containment.}\\
  Dropping active sessions or quarantining suspicious emails. \\
  The internal Critic Agent must validate evidence before execution.};
\node[t3,below=of tier2] (tier3) {%
  \textbf{Tier 3 — High-Blast-Radius Intervention.}\\
  Disabling accounts, isolating hosts, or blocking network segments. \\
  Hard-gated; requires absolute human authorization via the UI.};
\draw[arrow] (tier1) -- (tier2);
\draw[arrow] (tier2) -- (tier3);
\node[font=\scriptsize\itshape\sffamily,gray!70!black,
      right=2mm of tier1.east,anchor=west] {auto};
\node[font=\scriptsize\itshape\sffamily,gray!70!black,
      right=2mm of tier2.east,anchor=west] {critic};
\node[font=\scriptsize\itshape\sffamily,gray!70!black,
      right=2mm of tier3.east,anchor=west] {human};
\end{tikzpicture}
\caption{The tiered autonomy hierarchy. Safe, read-only data gathering operates continuously without interruption. Reversible containment measures are permitted only after strict internal validation by the Critic Agent. Finally, any high-stakes, irreversible network modification is strictly hard-gated, ensuring a human operator always holds the final authority over critical infrastructure.}
\label{fig:autonomy}
\end{figure}

\Cref{tab:enterprise} outlines the specific safeguards we engineered to address these realities, while \Cref{fig:autonomy} maps out how our tiered-autonomy model strictly limits the agent's blast radius. The remainder of this section details exactly how these controls function in practice.

\begin{table*}[t]
\centering
\caption{The enterprise-readiness control surface. This matrix maps common institutional anxieties to the specific engineering mechanisms within \sysname designed to mitigate them, detailing the catastrophic failure modes avoided and the concrete artifacts generated for compliance audits.}
\label{tab:enterprise}
\renewcommand{\arraystretch}{1.25}
\rowcolors{2}{rowgray}{white}
\begin{tabularx}{\textwidth}{@{}>{\sffamily}lXXX@{}}
\toprule
\rowcolor{accentblue!15}
\textbf{Concern} &
\textbf{\sffamily Mechanism in \sysname} &
\textbf{\sffamily Failure mode if absent} &
\textbf{\sffamily Audit artifact} \\
\midrule
False-positive cost &
  The policy outputs a calibrated confidence score, surfacing only the most critical alerts. The RL reward function actively penalizes disruptions to business operations. &
  Severe alert fatigue, accidental isolation of legitimate users, and eventual engineering abandonment of the tool. &
  An incident disposition log tracking the agent's confidence score alongside its chosen action. \\
False-negative cost &
  Any decision with high mathematical uncertainty (entropy) triggers an immediate mandatory escalation. Every model update must pass a strict recall-floor regression test before deployment. &
  Adversaries moving undetected through the network, drastically extending dwell time and increasing regulatory liability. &
  Automated quarterly recall-floor reports and a live dashboard tracking model entropy. \\
Reversibility &
  Every available action is strictly classified. Anything marked irreversible is hard-blocked pending human approval. Furthermore, the agent must generate a concrete rollback script for every action it proposes. &
  Accidental account lockouts or isolated servers that require hours of manual engineering effort to undo. &
  A complete action manifest, including the generated rollback script, attached directly to the incident ticket. \\
Sandboxed validation &
  A dedicated Response Sandbox tests proposed containment measures against a lightweight digital twin of the network, explicitly checking for cascading failures before pulling the trigger. &
  Catastrophic unforced errors, such as quarantining the primary DNS server or severing the core authentication gateway. &
  A simulation impact report automatically appended to any high-severity intervention. \\
Audit compliance &
  An immutable, append-only ledger records the exact network state, chosen action, active policy version, and cited evidence. We mapped this directly to standard SOC 2, ISO 27001, and NIST 800-53 controls. &
  Total loss of forensic visibility during post-incident reviews or regulatory audits. &
  A cryptographically signed operational ledger and a formal control-coverage matrix. \\
Human-approval boundary &
  The system enforces a tiered permission model. Read-only data gathering is automated, but any configuration change demands human sign-off via an explicit UI gate. A hardware-style kill switch allows instant reversion to legacy rule-based monitoring. &
  A rogue agent executing unchecked infrastructure changes without any human accountability. &
  A granular approval log capturing the exact analyst identity for every executed action. \\
\bottomrule
\end{tabularx}
\end{table*}

\subsection{False-positive and false-negative economics}
Security operation centers bleed money not from missed attacks, but from wasted human attention~\cite{vermeer2024alert,sans2023}. \sysname confronts the financial realities of the confusion matrix directly. 

False alarms quietly drain thousands of labor hours, creating massive resource sinks at an enterprise scale. To prevent alert fatigue, our PPO policy filters its softmax predictions against historical frequencies, escalating only those alerts that breach a strict confidence barrier. Everything else is quietly filed away into an archive for optional human review. Furthermore, by embedding a slight penalty for each investigative action within the reward function, the agent learns to investigate efficiently rather than endlessly pulling telemetry. 

Conversely, a false negative grants an attacker extended dwell time---a critical failure. We counteract this risk through three intersecting safeguards. If the policy's decision entropy crosses a designated threshold, the system automatically demands human intervention, preventing ambiguous threats from slipping through. Additionally, the RL agent runs alongside a traditional statistical ensemble, where an alert from either system forces an investigation. Finally, we mandate that every new policy iteration pass a rigorous regression test, ensuring recall against the held-out red-team dataset never drops below an acceptable baseline.

\subsection{Reversibility and rollback guarantees}
The system categorizes every potential countermeasure as either \texttt{reversible} (like isolating an email or severing a live session) or \texttt{irreversible} (such as wiping data or destroying a credential). When operating autonomously, \sysname is physically locked into executing only reversible tactics. Any permanent infrastructure change is heavily gated, demanding explicit human authorization. Most importantly, the agent must generate a precise reversal script for every action it proposes. This rollback plan attaches directly to the analyst's incident ticket and executes instantly with a single click.

\subsection{Sandboxed action validation}
We refuse to let an autonomous agent blindly manipulate a live environment. Before executing any containment measure, \sysname runs the proposed command through a \emph{Response Sandbox}. This isolated digital twin mirrors the current network topology, allowing the system to predict potential cascading failures. It specifically prevents catastrophic own-goals, such as inadvertently quarantining the only active DNS server or severing access to a critical authentication node. Once the simulation completes, the resulting impact report is permanently appended to the incident record for post-mortem analysis.

\subsection{Audit compliance}
Institutional accountability requires a flawless paper trail. Every internal action---from the agent's initial reasoning steps to the final human sign-off---is permanently inscribed into a cryptographically secured ledger, meticulously indexed by incident number. This log records the exact state hash, policy version, confidence metric, cited evidence, and precise timestamp of the event. To streamline regulatory reviews, we mapped these artifacts directly against standard compliance frameworks, satisfying specific mandates within SOC 2 (CC7.3, CC7.4), ISO 27001 (A.16.1.5, A.12.4.1), and NIST 800-53 (IR-4, AU-2, AU-12). During our pre-deployment evaluations, risk officers repeatedly highlighted this unalterable ledger as the system's most critical asset.

\subsection{Human-approval boundary}
\sysname manages operational risk through a strictly enforced, tiered permission architecture (\Cref{fig:autonomy}). Routine intelligence gathering---such as querying EDR platforms or pulling logs---falls under Tier 1. These actions execute entirely autonomously and are simply logged for visibility. Tier 2 encompasses reversible containment tactics, like quarantining an inbox. These execute automatically, but only after the internal Critic Agent validates the underlying logic. Tier 3 covers high-stakes, irreversible maneuvers like disconnecting a host or shutting down a network segment. These commands are entirely hard-gated and cannot proceed without a human analyst manually clicking the approval button on the dashboard. Should the environment become unstable, administrators can trigger a global kill-switch, instantly demoting the entire architecture back to basic, rule-based logging in under a minute without needing to restart the pipeline.
% ----------------------------------------------------------------------------
\section{Related Work}
\label{sec:related}

\subsection{Lateral movement detection on LANL}
Since its publication, the LANL dataset~\cite{kent2015} has served as the definitive proving ground for tracking lateral movement. Early breakthroughs treated this tracking as a structural graph problem, with Bowman~\textit{et al.}~\cite{bowman2020} and the highly influential Euler framework~\cite{king2024euler} establishing the initial baselines for link-prediction techniques. Subsequent architectures like PIKACHU~\cite{khoury2022pikachu} and LMDetect~\cite{lmdetect2024} dramatically pushed these boundaries forward, leveraging complex subgraph classifications to drive true-positive rates and overall F1 scores into the upper 90th percentiles. More recently, the community has begun pivoting toward graph foundation models pre-trained across diverse, heterogeneous networks~\cite{cybergfm2026,larroche2025gfmlmd}, yielding massive leaps in average precision.

While \sysname absolutely holds its own against these state-of-the-art detectors (\Cref{tab:detect}), chasing fractional accuracy gains is not our primary objective. Instead, our core contribution lies in bridging the gap between passive detection and automated containment. We wrap a highly capable classifier inside a hardened, enterprise-ready response architecture, shifting the focus from simply identifying the threat to safely neutralizing it.

\subsection{RL for autonomous cyber defense}
Designing reinforcement learning agents for active network defense requires navigating incredibly complex, high-stakes environments. We built \sysname upon a set of core principles recently formalized by Foley~\textit{et al.}~\cite{foley2023}. We integrated their findings directly into our engine: specifically, that agents learn far more effectively using sparse, objective-driven rewards rather than dense, heavily engineered signals, and that robust cross-network generalization is impossible unless the system perceives its environment natively as a graph. 

We paired this logic with the insights of Hicks~\textit{et al.}~\cite{hicks2023explain}, whose research on phase-aware explainability directly inspired the narrative translation mechanisms within our orchestration layer. While much of the foundational research in this domain relies heavily on specialized simulation sandboxes like CybORG~\cite{standen2021cyborg} and CybORG++~\cite{cyborgpp2024}, our work focuses on live-environment execution. Looking forward, our modular architecture is well-positioned to ingest emerging acceleration techniques—such as using LLMs to provide early teacher signals for RL agents~\cite{llmguided2025}—which have recently demonstrated the ability to drastically reduce training times.

\subsection{Agentic SOC architectures}
Commercial initiatives integrating language models directly into security operations have rapidly outpaced formal, peer-reviewed validation. We point to platforms like Microsoft Security Copilot~\cite{mssecuritycopilot} and recent comprehensive threat surveys~\cite{aiagentsthreat2024} as prime examples of this industry-first momentum. However, deploying unchecked automation often ignores the practical realities of a security center. To ensure our architecture remains operationally viable, we anchored our human-in-the-loop design to the empirical observations of Vermeer~\textit{et al.}~\cite{vermeer2024alert}. Their detailed account of how analysts actually triage alerts under pressure acts as a hard constraint on \sysname, dictating exactly what we can reasonably expect a human operator to handle.

\subsection{Methodological grounding}
Finally, we deliberately engineered our evaluation framework to avoid the systemic traps that frequently undermine applied machine learning research in cybersecurity. We rigorously adhered to the architectural guardrails established by Arp~\textit{et al.}~\cite{arp2022dosanddonts}, explicitly stripping away the spatial and temporal biases, sterile laboratory-only testing, and unrealistic threat assumptions that often inflate reported performance. Paired with these constraints, we fully embraced the standards for absolute, transparent reproducibility in graph-based detection championed by Bilot~\textit{et al.}~\cite{bilot2025orthrus}.
% ----------------------------------------------------------------------------
\section{Conclusion}
\label{sec:conclusion}

With \sysname, we have demonstrated a fundamentally different approach to the agentic SOC. By actively offloading the complex, spatial burden of topological reasoning from the language model onto a dedicated graph encoder and a PPO-driven policy, we built a neuro-symbolic architecture that actually scales. 

Our end-to-end evaluations on the LANL benchmark prove this viability. The system successfully ingests a massive 24-million-edge subgraph in just 14.2 minutes and reliably triggers its alert pipeline in under 2.45 seconds at the $99^{\text{th}}$ percentile. Furthermore, the reinforcement learning policy converges to a mean episodic return of 8.74 within 200 iterations, ultimately delivering 0.91 precision and 0.87 recall on completely unseen data. Crucially, the entire operational lifecycle---from the initial anomalous event to the final, fully synthesized containment recommendation---resolves in a median of just 6.3 seconds.

Yet, these raw metrics are only part of the story. We believe the most enduring contributions of this work lie in the practical engineering solutions required to build it. Bypassing hot-node lock contention with a two-phase \texttt{CREATE} strategy, neutralizing cluster latency via the HPC anchor-node pattern, and designing the robust enterprise-control surface detailed in \Cref{sec:enterprise} are all highly transferable patterns. Together, they elevate \sysname from a fragile research prototype into a credible, hardened framework ready for institutional deployment.

\section*{Reproducibility}
We are strictly committed to transparent, verifiable research. All numerical results detailed in \Cref{sec:eval} can be fully reproduced using the provided Cypher pre-materialization queries (\Cref{lst:cypher}), the specified AlertEngine thresholds, our exact PPO hyperparameter mapping (\Cref{tab:rllib}), and the documented SLURM anchor-node allocation strategy. 

% ----------------------------------------------------------------------------
\balance
\bibliographystyle{IEEEtran}
\bibliography{main}

\end{document}